\documentclass[12pt]{article}
\usepackage{amsmath}
\usepackage{environ}
\usepackage{etoolbox}
\usepackage{hyperref}
\hypersetup{colorlinks,linkcolor={blue},citecolor={blue},urlcolor={blue}}
\usepackage[pdftex]{graphicx}
\usepackage{graphics}
\usepackage[version=4]{mhchem}
\usepackage{pict2e}
\usepackage{mathtools, nccmath}
\usepackage{diffcoeff}
\usepackage{setspace}
\usepackage{etoc}
\usepackage{cite}
\usepackage{float}
\usepackage{lipsum}
\usepackage{blindtext}
\usepackage[table,xcdraw]{xcolor}
\usepackage{multirow}
\usepackage{booktabs}
\usepackage{siunitx}
\usepackage{booktabs}
\usepackage{outline}
\usepackage{avant} 
\usepackage{multirow}
 \usepackage[table,xcdraw]{xcolor}

\usepackage{array}
\usepackage{rotating}
\usepackage{tikz}
\usepackage{lscape}
\usepackage{subcaption}
\usepackage[toc,page]{appendix}
\usepackage{enumitem}

\title{\textbf{Muon ($g-2$) and W-boson mass Anomaly in a Model Based on $Z_4$ Symmetry with Vector like Fermion}}

\author {Simran Arora\thanks{ 009simranarora@gmail.com}, Monal Kashav\thanks{monalkashav@gmail.com}, Surender Verma\thanks{s\_7verma@hpcu.ac.in, Corresponding Author} and B. C. Chauhan\thanks{bcawake@hpcu.ac.in}}

\date{\textit{Department of Physics and Astronomical Science,\\Central University of Himachal Pradesh, Dharamshala 176215, INDIA.}}

\begin{document}
\maketitle

\begin{abstract}
\noindent The latest results of CDF-II collaboration show a discrepancy of $7\sigma$ with standard model expectations. There is, also, a $4.2\sigma$ discrepancy in the measurement of muon magnetic moment reported by Fermilab. We study the connection between neutrino masses, dark matter, muon ($g-2$) and W-boson mass anomaly within a single coherent framework based on $Z_{4}$ extension of the scotogenic model with vector like lepton (VLL). Neutrino masses are generated at one loop level. The inert doublet, also, provide a solution to W-boson mass anomaly through correction in oblique parameters $S$, $T$ and $U$. The coupling of VLL triplet $\psi_T$ to inert doublet $\eta$ provides positive contribution to muon anomalous magnetic moment. In the model, the VLL triplet provides a lepton portal to dark matter ($\eta_R^0$). The model predicts a lower bound $m_{ee}>0.025$ eV at 3$\sigma$, which is well within the sensitivity reach of the  $0\nu\beta\beta$ decay experiments. The model explains muon anomalous magnetic moment $\Delta a_\mu$ for $1.3<y_\psi<2.8$ and mass of DM candidate in the range $152\text{ GeV}<M_{\eta_{R}^{0}}<195\text{ GeV}$. The explanation of W-boson mass anomaly, further, constrain the mass of DM candidate, $M_{\eta_{R}^{0}}$,  in the range $154\text{ GeV}<M_{\eta_{R}^{0}}<174\text{ GeV}$.  
\end{abstract}
\section{Introduction}
The existence of dark matter (DM), non-zero neutrino masses and  matter-antimatter asymmetry suggest that the standard model (SM) of particle physics, despite its enormous triumphs, cannot be regarded as a conclusive hypothesis. Very recently, muon ($g-2$) Collaboration at Fermilab has reported the measurement of the  muon anomalous magnetic moment $ a_{\mu} \equiv (g_{\mu}-2)/2$ showing a $4.2\sigma$ discrepancy with SM prediction \cite{Muong-2:2021ojo}. The difference between the combined experimental result $a_{\mu}^{exp}$ = $116 592 061(41) \times 10^{-11}$ \cite{Muong-2:2021ojo} and SM prediction $a_{\mu}^{SM}$ = $116 591 810(43) \times 10^{-11}$ \cite{Aoyama:2020ynm} of muon anomalous magnetic moment is $\Delta a_{\mu} \equiv a_{\mu}^{exp}-a_{\mu}^{SM} = 251(59)\times 10^{-11}$ 
which is a sign of new physics (NP) beyond the SM. Several models have been proposed with an aim to accommodate muon ($g-2$). Although, there exist theoretical models based on discrete symmetries to resolve muon anomaly \cite{Abe:2017jqo,Hutauruk:2020xtk,Calibbi:2021qto,Saez:2021qta} but the most general framework include the extension of SM by $U(1)_{L_{\mu}-L_{\tau}}$ symmetry \cite{He:1990pn,He:1991qd}, wherein an additional gauge boson provide correction to muon magnetic moment at one loop \cite{Borah:2021mri,Biswas:2016yan, Zhou:2021vnf,Guo:2006qa,Dev:2017fdz, Majumdar:2020xws,Patra:2016shz}.

\noindent Within SM, the mass of W-boson is $M_{W}^{SM} = 80.354\pm 0.007$ GeV \cite{ParticleDataGroup:2020ssz}, however, CDF-II collaboration has recently published their results for the mass of W gauge boson $M_{W}$ = 80.4335 $\pm$ 0.0094 GeV \cite{CDF:2022hxs} which shows $7\sigma$ discrepancy with corresponding SM prediction\cite{ParticleDataGroup:2020ssz}. There have been many attempts to resolve this anomaly \cite{CentellesChulia:2022vpz,YaserAyazi:2022tbn,Afonin:2022cbi,Kawamura:2022fhm,Wang:2022dte,Batra:2022pej,Borah:2022zim,Borah:2022obi,Popov:2022ldh,Ghorbani:2022vtv}. The authors in Refs. \cite{Chakrabarty:2022voz,Chowdhury:2022dps,He:2022zjz,Dcruz:2022dao,Kim:2022xuo,Kim:2022hvh,Botella:2022rte,Zhou:2022cql,Baek:2022agi,Bhaskar:2022vgk,Han:2022juu} have attempted to provide a common explanation for both muon ($g-2$) and W-boson mass anomaly.

In the context of scotogenic models, $U(1)_{L_{\mu}-L_{\tau}}$ symmetry has been employed where new gauge boson emanating from the symmetry breaking explains the muon($g-2$) \cite{Borah:2021khc,Jana:2020joi,Baek:2015fea,Han:2019diw,Kang:2021jmi}. In such scenarios, the scotogenic fields ($\eta, N_{k}$) do not contribute to muon ($g-2$), explicitly.  Alternatively, in the present work, $Z_4$ symmetry is employed for the coupling of muon to the inert doublet $\eta$ and vector like lepton (VLL), $\psi_T$, explaining muon ($g-2$) in the framework of scotogenic model \cite{Ma:2006km}. The novelty of the framework lies in the fact that inert doublet ($\eta$) contributes to neutrino mass, dark matter, muon ($g-2$) and and W-boson anomaly explanation. With the addition of three right-handed neutrinos and an inert doublet, the scotogenic model offers a simultaneous explanation for dark matter and non-zero neutrino masses. A scalar singlet is added to the particle content of the model which breaks $Z_{4}$ to $Z_{2}$ symmetry. The inert doublet and right-handed neutrinos are odd under the unbroken $Z_{2}$ symmetry and stabilizes the lightest DM candidate that is real part of inert doublet in the model. The neutrino masses are generated at one loop level. In order to accommodate the muon ($g-2$), a VLL triplet is added which when couples to the scalar doublet results in chirally enhanced positive contribution to muon ($g-2$) and, also, provide a lepton portal to DM. In addition, the model can resolve W-boson mass anomaly through correction in oblique parameters $S$, $T$ and $U$ reflecting the NP contribution. 

\noindent The paper is organized as follows. In Section 2, we have discussed the basic structure and phenomenological consequences of the scotogenic model. In Section 3, we outline the numerical analysis and related discussion. Finally, in Section 4 we summarize our results.

\section{Scotogenic Model for Neutrino Masses}
We have extended the SM gauge symmetry by a $Z_{4}$ symmetry. In order to generate small neutrino masses, we have considered scotogenic model that extends SM with three right-handed neutrinos and an inert scalar doublet with an imposed $Z_{2}$ symmetry that, also, provides a solution to DM. The particle content of the model with corresponding charge assignments is given in Table \ref{tab1}.  Here, $L_{\alpha}$ and  $\alpha_{R}$  $(\alpha = e, \mu, \tau)$ are left-handed lepton doublets and right-handed charged leptons, respectively, $N_{k} (k=1,2,3)$ are right handed neutrino singlets, $\psi_{T}$ is VLL triplet, $H$ is SM Higgs, $S$ is scalar singlet and $\eta$ is inert scalar doublet.  

\begin{table}[t] \label{Tab1}
\centering
 \begin{tabular}{c c c c c c c c} 
 \hline\hline

 Symmetry Group & $L_{e}$, $L_{\mu }$, $L_{\tau }$ & $e_{R}$, $\mu_{R}$, $\tau_{R}$ &$N_{1}$, $N_{2}$, $N_{3}$ & $\psi_{T}$ & $H$ & $S$ & $\eta$ \\ [0.5ex] 
 \hline
 $SU(2)_{L}$ $\times$ $U(1)_{Y}$ & (2, -1/2) & (1, -1) & (1, 0)&(3, -1) & (2, 1/2) & (1, 0)&(2, 1/2) \\ 
$ Z_{4}$ &(1, i, -i) & (1, i, -i) & (1, i, -i)&i &1&-i&1 \\
$Z_{2}$ & + & + & -- & -- & +&+& -- \\ [1ex] 
 \hline
 \end{tabular}
  \caption{The field content and respective charge assignments of the model under $SU(2)_{L}\times U(1)_{Y}\times Z_{4}\times Z_{2}$.}
      \label{tab1}
\end{table}

The relevant terms in the Yukawa Lagrangian are

    \begin{eqnarray} \label{Ly}
    \nonumber
-\mathcal{L} &\supseteq&  \frac{M_{11}}{2} N_{1} N_{1} + M_{23} N_{2} N_{3} + y_{\eta 1} \bar{L}_{e} \tilde{\eta} N_{1} + y_{\eta 2} \bar{L}_{\mu} \tilde{\eta} N_{2 } + y_{\eta 3} \bar{L}_{\tau} \tilde{\eta} N_{3} \nonumber \\ &&
    +\:  y_{12} S  N_{1} N_{2} + y_{13} S^{*} N_{1} N_{3} +  y_{\psi} \eta^{\dagger}\bar {\psi}_{T,R}L_{\mu} + M_{\psi} \bar{\psi_{T}} \psi_{T} \nonumber \\ &&
    +\: y_{ e} \bar{L}_{e} H e_{R} + y_{\mu} \bar{L}_{\mu} H \mu_{R} + y_{\tau} \bar{L}_{\tau} H \tau_{R} + H.c.,
   \end{eqnarray}
with $\tilde{\eta}=i \sigma_{2}\eta^{*}$ and the scalar potential $V(H,S,\eta)$ is
  
   \begin{eqnarray}
    \nonumber
    V(H,S,\eta) & = & -\mu_{H}^{2}(H^{\dagger} H)+\lambda_{1}(H^{\dagger} H)^{2}-\mu_{S}^{2}(S^{\dagger}S)
     +\lambda_{S}(S^{\dagger
}S)^{2}+\lambda_{HS}(H^{\dagger} H)(S^{\dagger} S) \nonumber \\ && +\:\mu_{\eta}^{2}(\eta^{\dagger}\eta) 
+ \lambda_{2}(\eta^{\dagger}\eta)^{2} + \lambda_{3}(\eta^{\dagger}\eta)(H^{\dagger} H)\nonumber\\
     && \:+\lambda_{4}(\eta^{\dagger} H)(H^{\dagger} \eta) +\frac{\lambda_{5}}{2}[(H^{\dagger} \eta)^{2}+(\eta^\dagger H)^{2}]+
     \lambda_{\eta S}(\eta^{\dagger} \eta)(S^{\dagger} S) + H.c..
\end{eqnarray}

The SM gauge symmetry is broken by the neutral component of Higgs doublet $H$ while the $Z_{4}$ symmetry is broken softly by the non-zero vacuum expectation value ($vev$) of scalar singlet $S$ into the residual $Z_{2}$ symmetry. Also, we assume $\mu_{\eta}^2$ $>$ 0  so that $\eta$ do not acquire any $vev$. 
 Using Eqn. (\ref{Ly}), the charged lepton mass matrix $M_{l}$,  Dirac Yukawa matrix $y_{D}$ and right-handed neutrino mass matrix $M_{R}$ are given by

\begin{eqnarray*}
M_{l} = \frac{1}{\sqrt{2}}\begin{pmatrix}
   y_{e}v && 0 && 0 \\
   0 &&  y_{\mu}v && 0 \\
   0 && 0 &&  y_{\tau}v
   \end{pmatrix},
\end{eqnarray*}

\begin{eqnarray}
  y_{D} = 
   \begin{pmatrix}
   y_{\eta 1} && 0 && 0 \\
   0 &&  y_{\eta 2} && 0 \\
   0 && 0 &&  y_{\eta 3}
   \end{pmatrix}  , M_{R} =
    \begin{pmatrix}
   M_{11} && y_{12}v_{S}/\sqrt{2} && y_{13}v_{S}/\sqrt{2} \\
   y_{12}v_{S}/\sqrt{2} &&  0 && M_{23}e^{i\delta} \\
   y_{13}v_{S}/\sqrt{2} && M_{23}e^{i\delta} && 0
   \end{pmatrix},
\end{eqnarray}
where $v/\sqrt{2}$ and $v_{S}/\sqrt{2}$ are $vevs$ of the Higgs field $H$ and scalar field $S$, respectively, and $\delta$ is the phase remaining after redefinition of the fields.

\subsection{Neutrino Masses, Muon ($g-2$), W-boson Mass Anomaly and Dark Matter}
\begin{figure}[t]
    \centering
    \includegraphics[width=12cm]{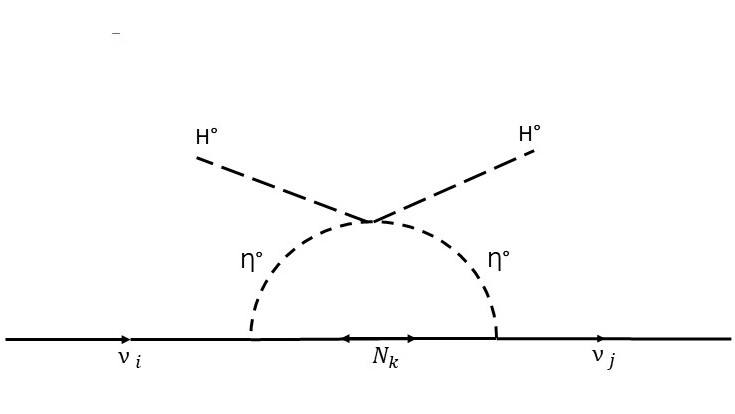}
    \caption{The diagram responsible for neutrino mass generation in scotogenic model at one loop level. }
    \label{fig:1}
\end{figure}

The neutrino masses are generated at one loop level (Fig. \ref{fig:1}) resulting in the light neutrino mass matrix given by\cite{Ma:2006km, Merle:2015ica}
 
 \begin{equation}\label{scot}
     M_{ij}^{\nu} = \sum_{k}{}\frac{y_{ik}y_{jk}M_{k}}{16\pi^{2}}\left[\frac{M_{\eta_{R}^{0}}^{2}}{M_{\eta_{R}^{0}}^{2}-M_{k}^{2}}\ln\frac{M_{\eta_{R}^{0}}^{2}}{M_{k}^{2}}-\frac{M_{\eta_{I}^{0}}^{2}}{M_{\eta_{I}^{0}}^{2}-M_{k}^{2}}\ln\frac{M_{\eta_{I}^{0}}^{2}}{M_{k}^{2}}\right],
 \end{equation}
  where $M_{k}$ is the mass of $k^{th}$ right-handed neutrino and $M_{\eta_{R}^{0},\eta_{I}^{0}}$ are the masses of real and imaginary parts of inert doublet $\eta$. The Yukawa couplings appearing in Eqn. (\ref{scot}) are derived from $y_D$ in the basis where $M_{R}$ is diagonal. 
 If the mass-squared difference between $\eta_{R}^{0}$ and $\eta_{I}^{0}$  i.e. ${M_{\eta_{R}^{0}}^{2}}-M_{\eta_{I}^{0}}^{2} = \lambda_{5}v^{2}$  $<<$  $M^{2}$ where $M^2 = (M_{\eta_{R}^{0}}^{2}+M_{\eta_{I}^{0}}^{2})/2$, then above expression reduces to 
 \begin{equation}\label{mmatrix}
    M_{ij}^{\nu}=\frac{\lambda_{5} v^{2}}{16 \pi^{2}}\sum_{k}{}\frac{y_{ik}y_{jk}M_{k}}{M^{2}-M_{k}^{2}}\left[1-\frac{M_{k}^{2}}{M^{2}-M_{k}^{2}}\ln\frac{M^{2}}{M_{k}^{2}}\right].
\end{equation}

The low energy effective neutrino mass matrix $M^\nu$ obtained using Eqn. (\ref{mmatrix}) can be diagonalized to ascertain model predictions for neutrino masses and mixing angles $viz.,$
\begin{equation}
    M^\nu=U M^\nu_d U^T,
\end{equation}
where $U$ is unitary matrix and $M^\nu_d=diag(m_1,m_2,m_3)$, $m_i$ are neutrino mass eigenvalues.  
In term of the elements of the diagonalizing matrix
\begin{eqnarray}
   U =
    \begin{pmatrix}
   U_{e1} && U_{e2} && U_{e3} \\
     U_{\mu 1} && U_{\mu 2} && U_{\mu 3} \\
    U_{\tau 1} && U_{\tau 2} && U_{\tau 3} 
   \end{pmatrix},
\end{eqnarray}
the neutrino mixing angles can be evaluated using

\begin{eqnarray}
    \sin^{2}\theta_{13} = |U_{e3}|^{2} ,\quad   \sin^{2}\theta_{23} = \frac{|U_{\mu 3}|^{2}}{1 - |U_{e3}|^{2}} ,\quad \sin^{2}\theta_{12} = \frac{|U_{e2}|^{2}}{1 - |U_{e3}|^{2}}.
\end{eqnarray}

\begin{figure} [t]
    \centering
    {{\includegraphics[width=5cm]{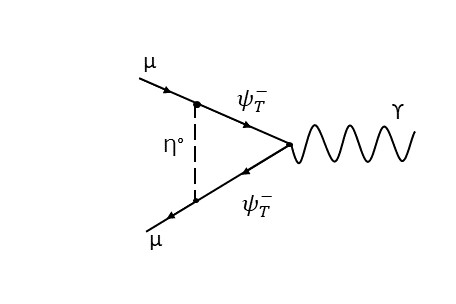} }}%
    \qquad
    {{\includegraphics[width=5cm]{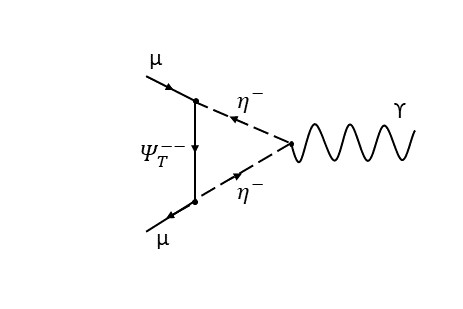}}}
    \qquad
    {{\includegraphics[width=5cm]{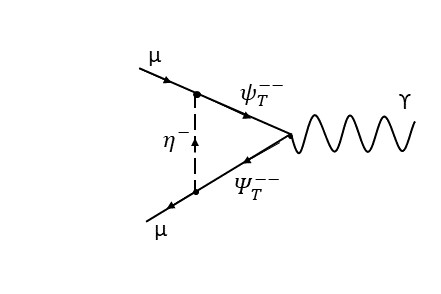}}}
    \caption{The diagrams responsible for positive contribution, from $\psi_{T}$ and $\eta$, to muon ($g-2$) at one loop level.}%
    \label{fig2}%
\end{figure}

\textbf{Contribution to Muon ($g-2$)}

In order to accommodate muon ($g-2$) in the model, we have added a VLL triplet $\psi_{T}$ with appropriate $Z_4$ charge so that it only couples to muon. This VLL $\psi_{T}$ alone gives negative contribution to muon ($g-2$)\cite{Freitas:2014pua}. Also, the charged component of scalar doublet $\eta$ has a negative contribution to muon ($g-2$). However, $\psi_{T}$ coupling with $\eta$, may results in chirally enhanced positive contribution to $\Delta a_{\mu}$ through $y_{\psi} \eta^{\dagger}\bar {\psi}_{T,R}L_{\mu}$ term in Eqn. (\ref{Ly}). The possible diagrams contributing to muon ($g-2$) are shown in Fig. \ref{fig2}. 

The contribution to muon anomalous magnetic moment is given by \cite{Freitas:2014pua}:
\begin{eqnarray} \label{au2}
 \Delta a_{\mu} = \frac{m_{\mu}^2 y_{\psi}^2}{32 \pi^2 M_{\eta}^2}[5 F_{FFS}(M_{\psi}^2/M_{\eta}^2) - 2 F_{SSF}(M_{\psi}^2/M_{\eta}^2)],
\end{eqnarray}
where 
$m_{\mu}$, $M_{\psi}$, $M_{\eta}$ are the masses of muon, VLL triplet $\psi_{T}$, inert scalar doublet $\eta$, respectively, and $y_\psi$ is coupling constant. Also,
\begin{eqnarray*}
  F_{FFS}(t) = \frac{1}{6(t-1)^4}[t^3 - 6t^2 + 3t + 2 + 6t \ln t],
\end{eqnarray*}
and
\begin{eqnarray}
   F_{SSF}(t) = \frac{1}{6(t-1)^4}[-2t^3 - 3t^2 + 6t - 1 + 6t^2 \ln t],
\end{eqnarray}
where $t=\frac{M_{\psi}^2}{M_{\eta}^2}$ and $M_{\eta}\approx M_{\eta_{R}^{0}}$.

\textbf{W-boson Mass Correction from the Scalar Doublet}

In scotogenic model, the scalar doublet $\eta$ provides correction to W-boson mass. The NP contribution is parameterized in terms of three self-energy parameters called \textit{``oblique parameters"} \textit{viz.} $S$, $T$ and $U$ \cite{Peskin:1991sw}. The correction to W-boson mass in terms of these parameters is given by\cite{Peskin:1991sw}

\begin{equation}\label{Mw}
    M_{W} = M_{W}^{SM}\left[1-\frac{{\alpha}_{em}}{4(c_{W}^{2}-s_{W}^{2})}(S-1.55T-1.24U)\right]
\end{equation}
where $M_{W}^{SM}$ is mass of W-boson in the SM, $\alpha _{em}$ is the electromagnetic fine structure constant, $c_{w}$ and $s_{w}$ are the $cos$ and $sin$ of Weinberg angle, respectively.
The value of the oblique parameters, including the new CDF-II W-boson mass result \cite{Lu:2022bgw} is 
\begin{equation}
S = 0.06 \pm 0.10, \quad T = 0.11 \pm 0.12 \quad \text{and} \quad U=0.14 \pm 0.09,
\end{equation}
with the correlation coefficients
\begin{equation}
\rho_{ST} = 0.90, \quad \rho_{SU}=-0.59 \quad \text{and} \quad \rho_{TU} = -0.85.
\end{equation}
With vanishing $U$, the values of $S$ and $T$ parameters are found to be \cite{Lu:2022bgw}
\begin{equation} \label{st}
  S=0.15 \pm 0.08  \quad \text{and} \quad T = 0.27 \pm 0.06 \quad \text{with correlation coefficient} \quad \rho_{ST}=0.93.
\end{equation}

With in the paradigm of scotogenic model, the main contribution to oblique parameters comes from the scalar doublet $\eta$. The $S$, $T$ and $U$ contributions from the scalar doublet $\eta$ have been derived assuming degenerate $CP$ even and $CP$ odd scalars ($M_{\eta_{R}^{0}}$ $\simeq$ $M_{\eta_{I}^{0}}$) since $\lambda_{5}$ is considered to be small. The $S$, $T$ and $U$ corrections are given by \cite{Zhang:2006vt}

\begin{eqnarray*}
  &&S \simeq\frac{1}{12\pi}\log\left(\frac{M^{2}}{M_{\eta^{+}}^{2}}\right),
   \end{eqnarray*}
     
  \begin{eqnarray}\label{T}
    &&T \simeq \frac{2\sqrt{2}G_{F}}{(4\pi)^{2}\alpha_{em}}\left[\frac{M^{2}+M_{\eta^{+}}^{2}}{2} - \frac{M^{2}M_{\eta^{+}}^{2}}{M_{\eta^{+}}^{2} - M^{2}}\log\left(\frac{M_{\eta^{+}}^{2}}{M^{2}}\right)\right],
  \end{eqnarray}

\begin{eqnarray*} 
 &&U \simeq \frac{1}{12\pi}\left[- \frac{5 M_{\eta^{+}}^{4}-22 M_{\eta^{+}}^{2}M^{2}+5 M^{4}}{3(M_{\eta^{+}}^{2}-M^{2})^{2}} +
 \frac{(M_{\eta^{+}}^{2}+M^{2})(M_{\eta^{+}}^{4}-4M_{\eta^{+}}^{2}M^{2}+ M^{4})}{(M_{\eta^{+}}^{2}-M^{2})^{3}} \log \left(\frac{M_{\eta^{+}}^{2}}{M^{2}}\right)\right],
\end{eqnarray*}
where $M^2 = (M_{\eta_{R}^{0}}^{2}+M_{\eta_{I}^{0}}^{2})/2$ and $M_{\eta^{+}}$ is the mass of charged component of inert doublet $\eta$.

\section{Numerical Analysis and Discussion}
\begin{figure} [t]%
    \centering
    {{\includegraphics[width=7cm]{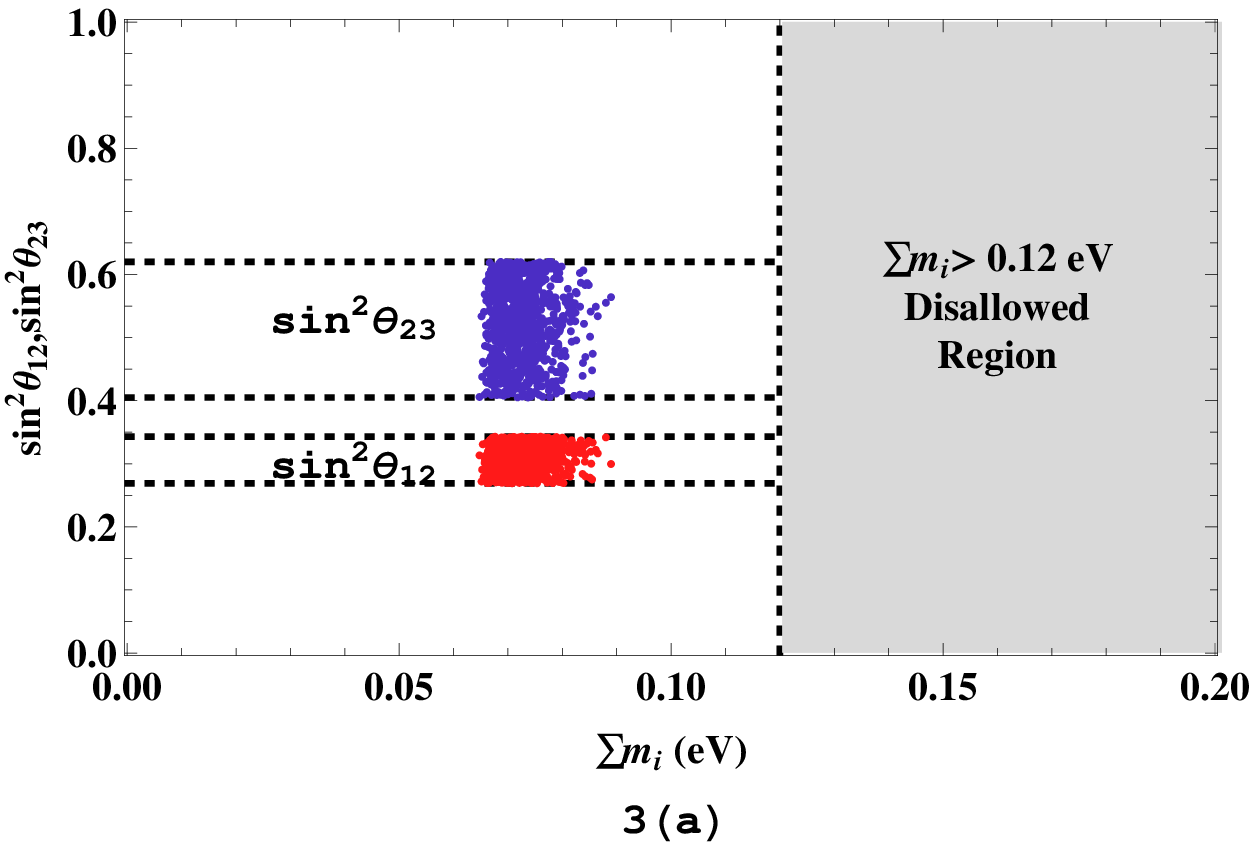} }}%
    \qquad
    {{\includegraphics[width=7cm]{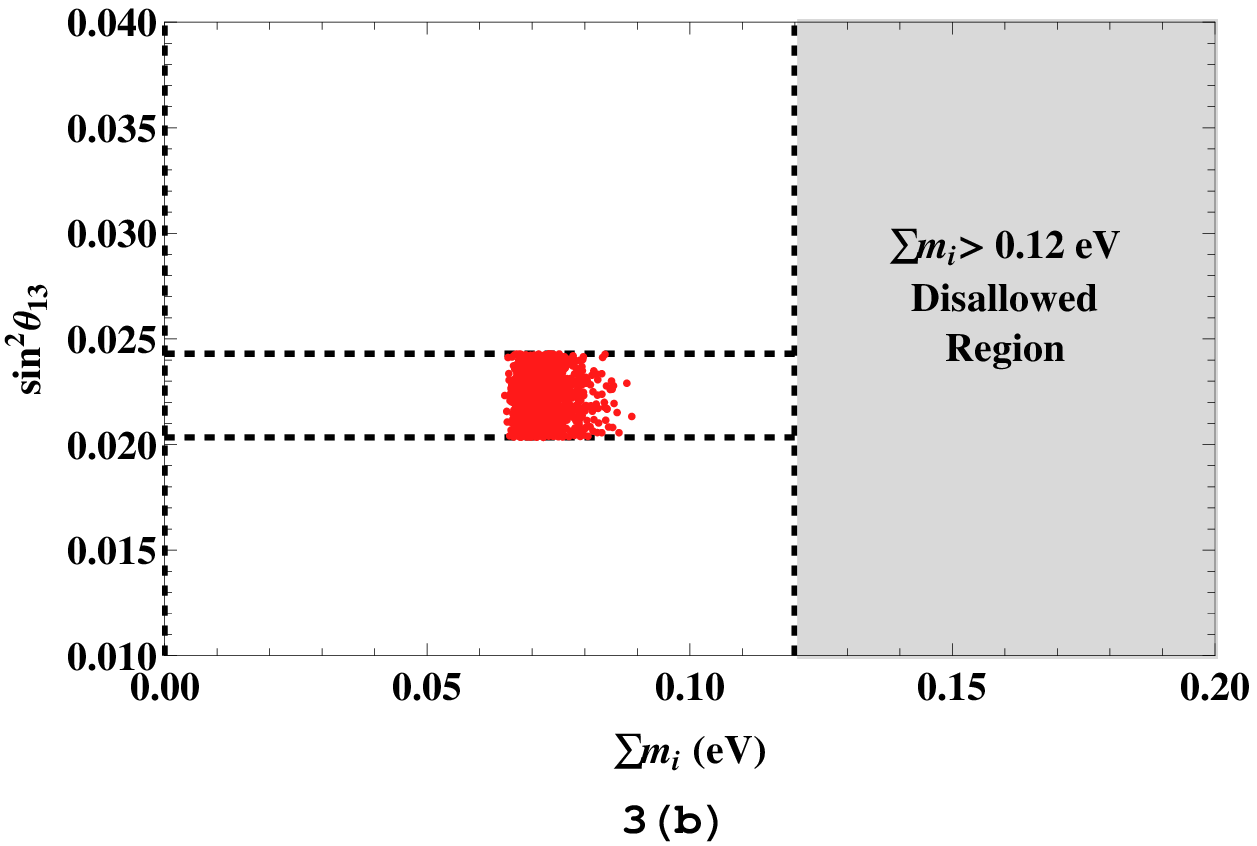}} }%
 \caption{The correlation of $\sin^2\theta_{12},\sin^2\theta_{23}$ and $\sin^2\theta_{13}$ with sum of neutrino masses $\Sigma m_{i}$. The horizontal lines are allowed $3\sigma$ ranges of mixing angles \cite{Esteban:2020cvm} and the grey shaded region is disallowed by cosmological bound on sum of neutrino masses \cite{Giusarma:2016phn,Planck:2018vyg}.}%
    \label{fig3}%
\end{figure}

In order to constrain the allowed parameter space and investigate neutrino phenomenology, the model predictions for neutrino mass-squared differences $(\Delta m_{21}^{2}, \Delta m_{31}^{2})$ and mixing angles $(\theta_{13}, \theta_{23},\theta_{12})$ are compared with 3$\sigma$ ranges of the experimental data on neutrino masses and mixings \textit{viz.,}  \cite{Esteban:2020cvm} 
 \begin{equation*}
    \sin^{2} \theta_{13} = (0.02034-0.02430) ,\quad  \sin^{2} \theta_{23} = (0.407 - 0.620) ,\quad \sin^{2} \theta_{12} = (0.269 - 0.343) ,
    \end{equation*}
    \begin{equation}\label{ndata}
  \Delta m^{2}_{31} = (2.431 - 2.599) \times 10^{-3} \text{eV}^{2} , \quad \Delta m^{2}_{21} = (6.82 - 8.04) \times 10^{-5} \text{eV}^{2}.
    \end{equation}
The model predictions for neutrino masses and mixing angles are obtained by numerically diagonalising the low energy effective neutrino mass matrix in Eqn. (\ref{mmatrix}). The free parameters of the model are varied randomly in ranges given in Table \ref{tab2}. The model predicts neutrino mixing angles and mass-squared differences within their $3\sigma$ range. For the sake of completeness, we have given the correlation plots depicting the allowed parameter space of the model. Fig. \ref{fig3}(a) shows the variation of $\sin^{2} \theta_{12}$ and $\sin^{2} \theta_{23}$ with sum of active neutrino masses $\Sigma m_i$ whereas Fig {\ref{fig3}}(b) shows the variation of $\sin^{2} \theta_{13}$ with sum of active neutrino masses $\Sigma m_i$.

Also, information about the $CP$ violation is encoded in $CP$ rephasing invariants $J_{CP}$, $I_1$ and $I_2$. The Jarlskog $CP$ invariant $J_{CP}$ is given by \cite{Krastev:1988yu,Jarlskog:1985ht} 

\begin{equation}
    \text{$J_{CP}$} = \text{Im}[U_{e1}U_{\mu 2}U_{e2}^{*}U_{\mu 1}^{*}],
\end{equation}
while the other two $CP$ invariants $I_{1}, I_{2}$ related to Majorana phases can be written as

\begin{align}
 I_{1} = \text{Im}[U_{e1}^{*}U_{e2}] ,\quad \quad I_{2} = \text{Im}[U_{e1}^{*}U_{e3}].
\end{align}

\begin{figure} [t]%
    \centering
    {{\includegraphics[width=7cm]{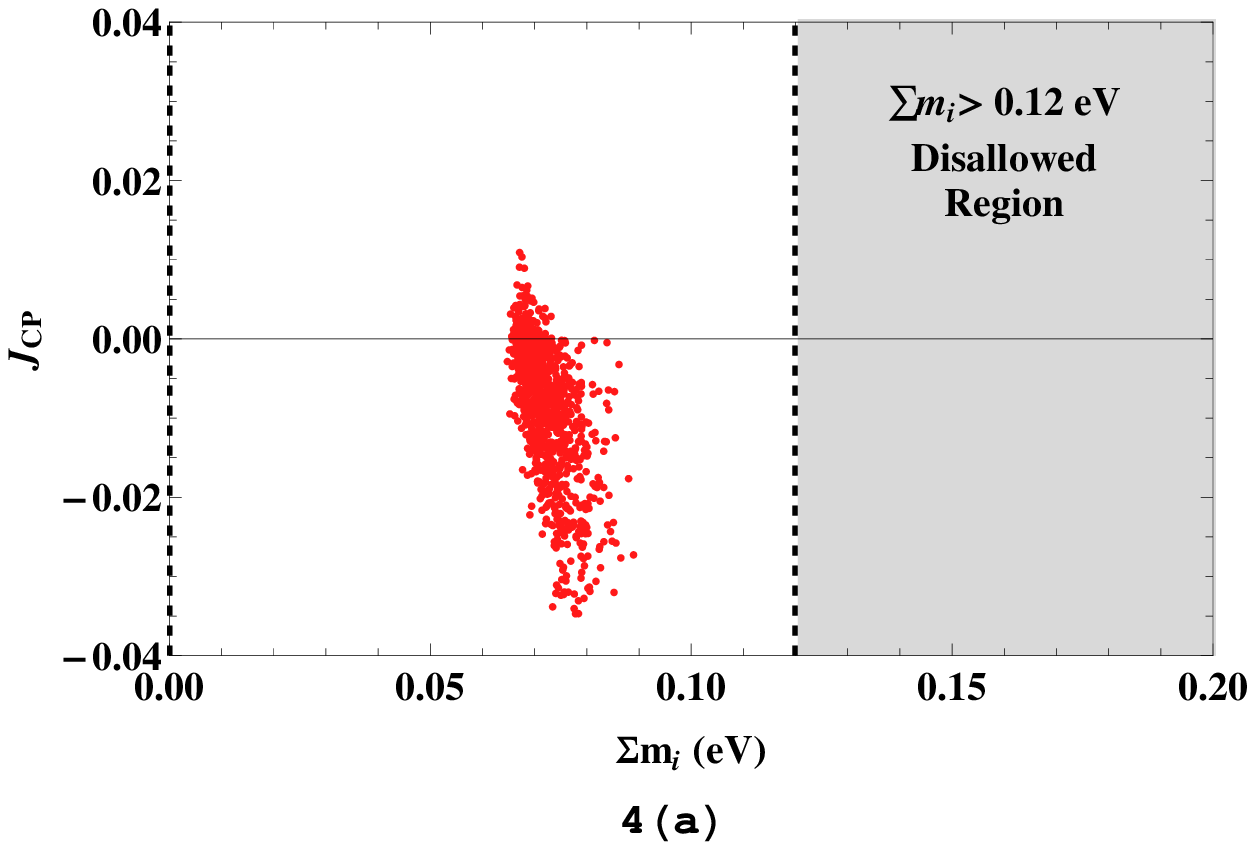}}}%
    \qquad
{{\includegraphics[width=7cm]{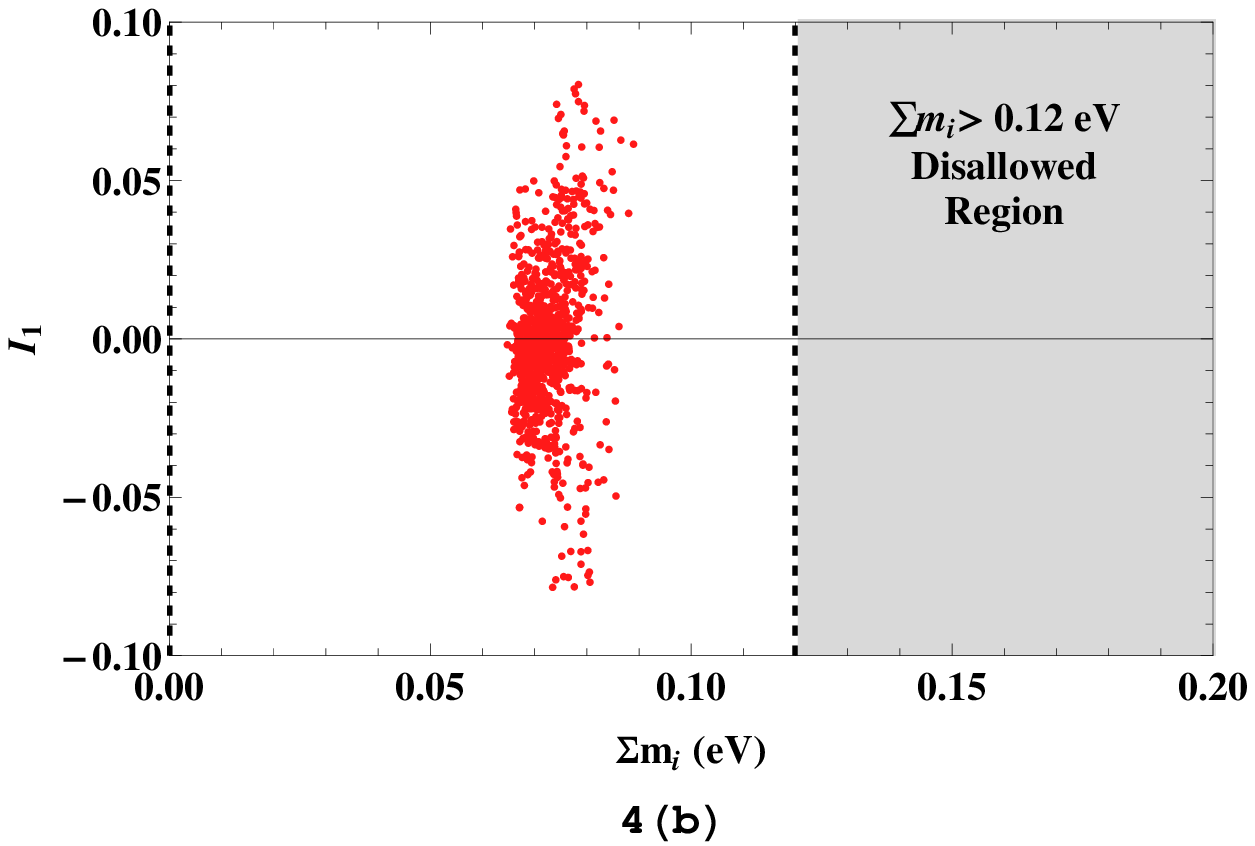} }}%
    \qquad
    {{\includegraphics[width=7cm]{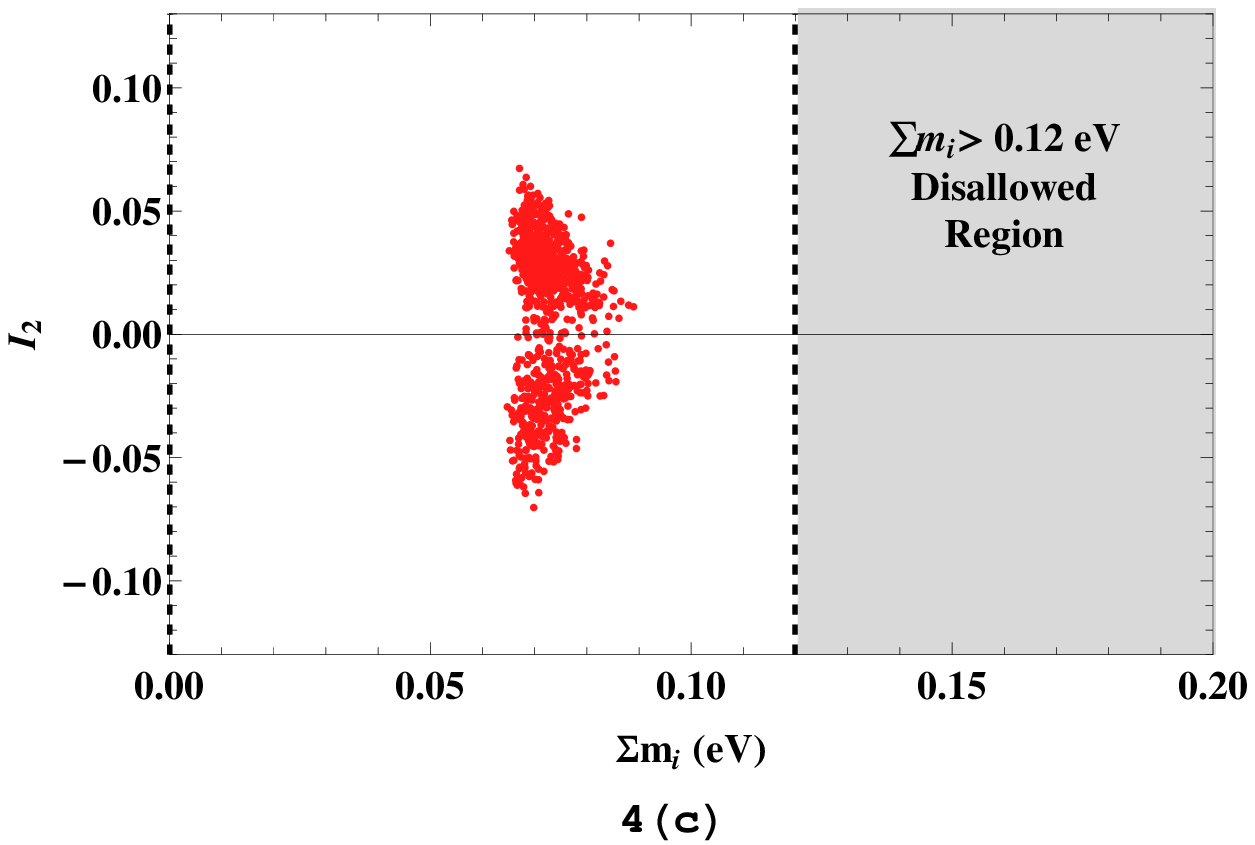} }}%
    \caption{The correlations of $CP$ rephasing invariants $J_{CP},I_{1}$ and $I_{2}$ with sum of neutrino masses $\Sigma m_{i}$. The grey shaded region is disallowed by the cosmological bound on sum of neutrino masses \cite{Giusarma:2016phn,Planck:2018vyg}.}%
  \label{fig4}%
\end{figure}

Fig. \ref{fig4}(a) shows the variation of Jarlskog $CP$ invariant with sum of active neutrino masses $\Sigma m_{i}$ and Figs. \ref{fig4}(b) and \ref{fig4}(c) show the correlation of $I_1$ and $I_2$ with sum of active neutrino masses $\Sigma m_{i}$. The model predicts both $CP$ conserving and violating solutions. 

\begin{figure} [t]%
    \centering
    {{\includegraphics[width=7cm]{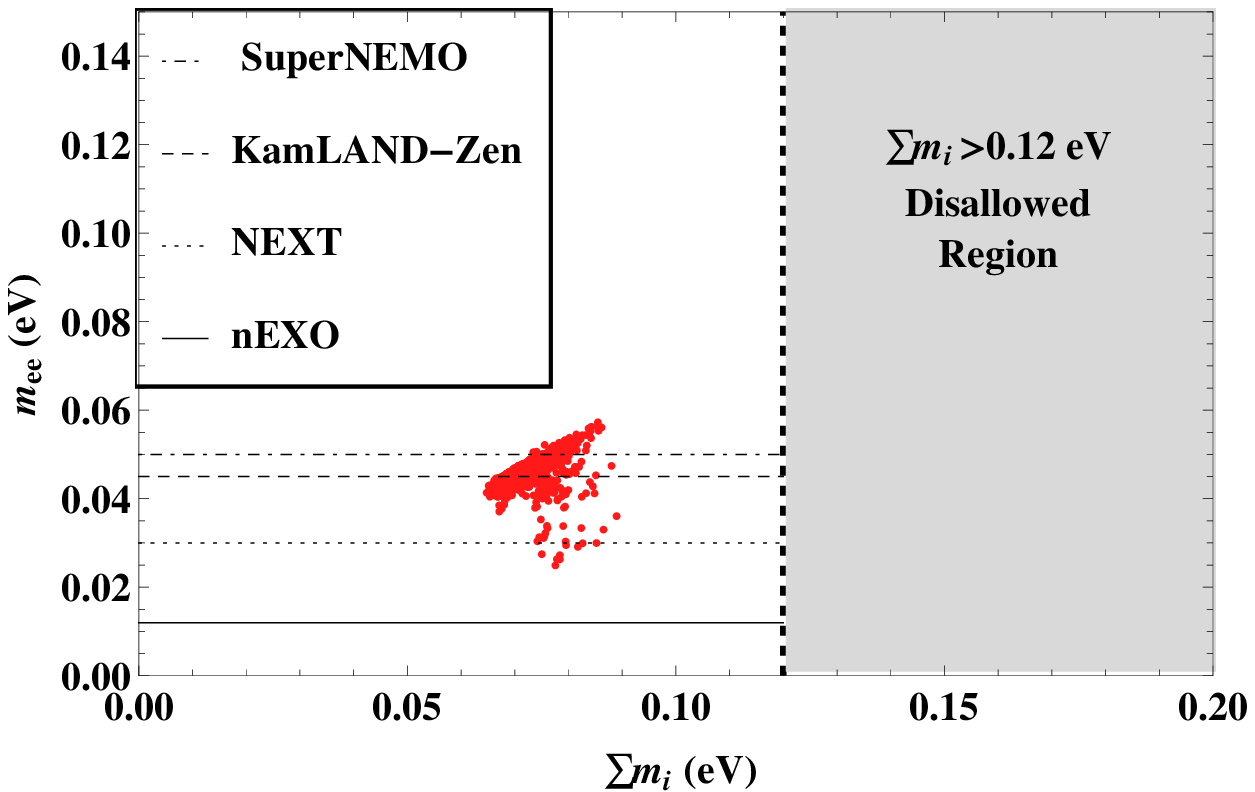} }}%
   \caption{The correlation of $m_{ee}$ with sum of neutrino masses $\Sigma m_{i}$. The sensitivity reach of various $0\nu\beta\beta$ decay experiments are shown as the horizontal lines. The grey shaded region is disallowed by the cosmological bound on sum of neutrino masses \cite{Giusarma:2016phn,Planck:2018vyg}.}%
  \label{fig5}%
\end{figure}

There is, also, a longstanding question in particle physics about the exact nature of neutrinos. Neutrinoless double beta decay ($0\nu\beta\beta$) experiment can confirm the Majorana nature of neutrino. The amplitude of this process is proportional to the (1,1) element of neutrino mass matrix for which the general expression is given by

\begin{equation}
     m_{ee} \equiv M^{\nu}_{11} = \left|\sum_{i=1}^{3} U_{ei}^{2}m_{i}\right|
\end{equation}
 The correlation of $m_{ee}$ with sum of active neutrino masses $\Sigma m_{i}$ is shown in Fig. \ref{fig5}. The sensitivities of $0\nu\beta\beta$ decay experiments such as nEXO\cite{Licciardi:2017oqg}, NEXT\cite{NEXT:2009vsd,NEXT:2013wsz}, KamLand-Zen\cite{KamLAND-Zen:2016pfg} and SuperNEMO\cite{Barabash:2011row} are, also, shown in the Fig. \ref{fig5}. The model predicts a lower bound $m_{ee}>0.025$ eV at 3$\sigma$, which is well within the sensitivity reach of the  $0\nu\beta\beta$ decay experiments.

\begin{table}[t]
    \centering
    \begin{tabular}{ll}
    \hline
    \hline
       Parameters  &  Range \\
       \hline
       
        $(M_{11} ,M_{23})$ & (10, 100) GeV \\ 
          ($x$, $y$) &  (10, 100) GeV \\ 
         $ \delta^{\circ}$ & (0, 360) \\ 
             $y_{\eta k}$  & $(10^{-2}, 10^{-1})$  \\
                $\mu_{\eta}$ & (10, 100) GeV \\ 
                 $v_{S}$ & (100, 400) GeV\\
                 $\lambda_{\eta S}$ & $(10^{-2}, 10^{-1})$ \\ 
                $ \lambda_{3}$ & $(10^{-2}, 10^{-1})$ \\ 
                 $\lambda_{4}$ & $(10^{-1},1)$ \\ 
                   $ \lambda_{5}$ & $(10^{-5}, 5\times 10^{-5})$\\ 
                     $ M_{k}$&$ (1 \times 10^{7}, 9 \times 10^{7})$ GeV  \\ 
                      \hline
    \end{tabular}
    \caption{The ranges of parameters used in the numerical analysis ($k=1,2,3$, $x\equiv \frac{y_{12} v_{S}}{\sqrt{2}}$ and $y\equiv \frac{y_{13} v_{S}}{\sqrt{2}}$).}
    \label{tab2}
\end{table}

\subsection*{Muon ($g-2$)}

\begin{figure}[t]%
    \centering
    \includegraphics[width=7cm]{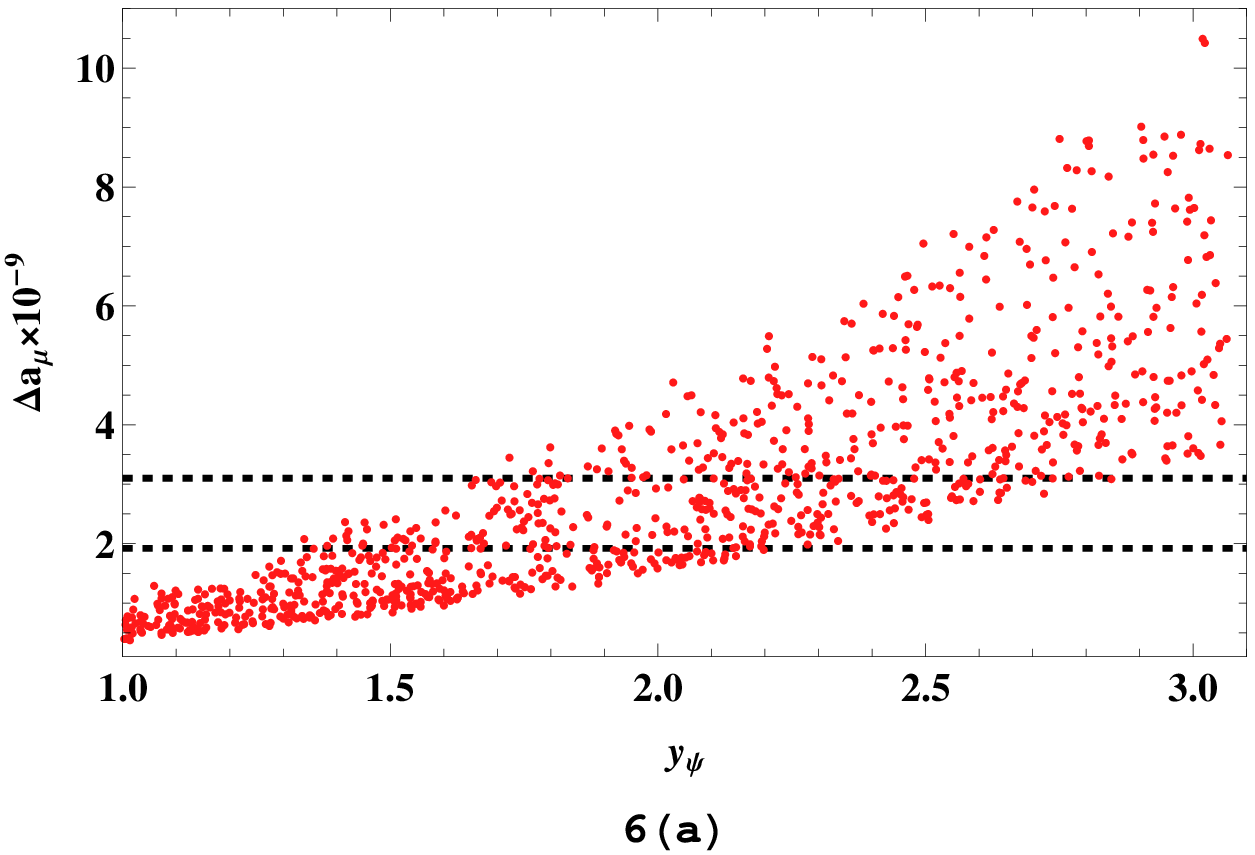}
    \qquad
     {{\includegraphics[width=7cm]{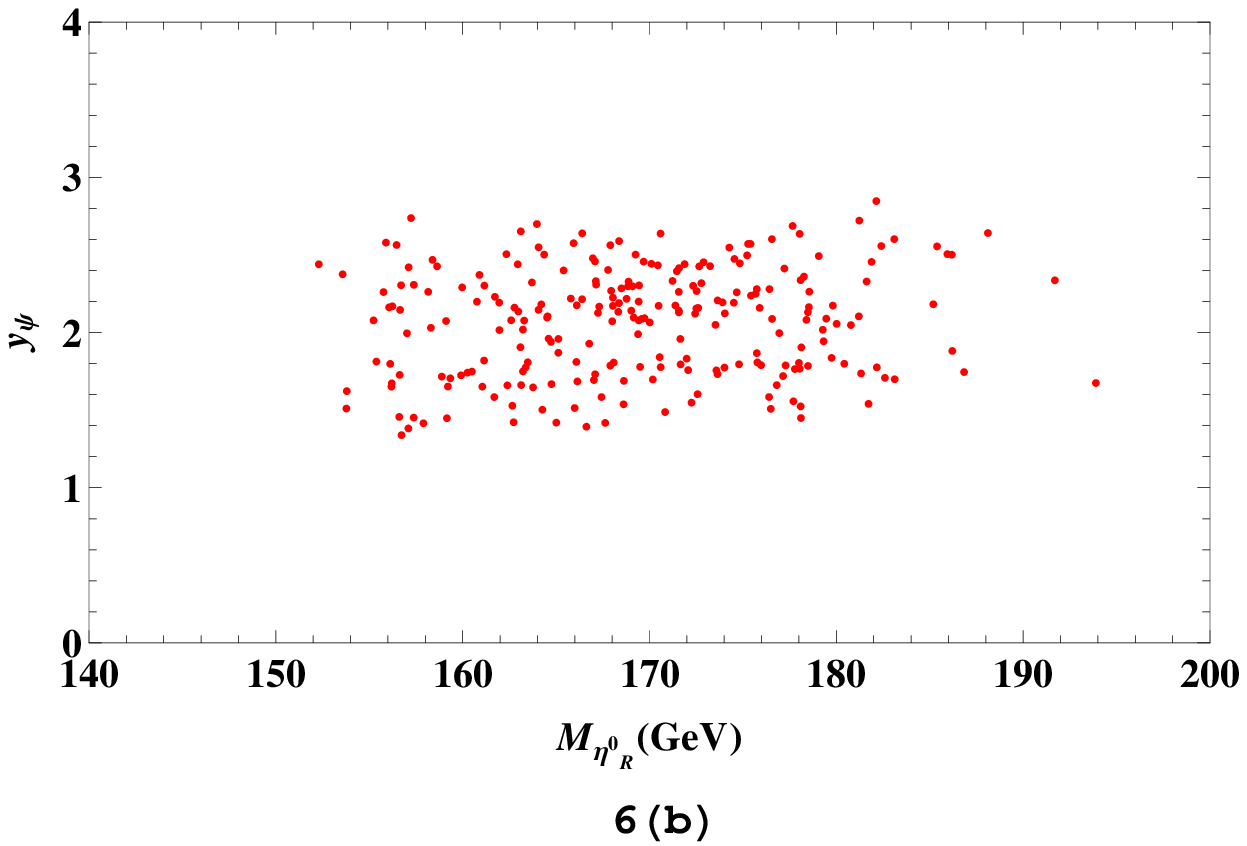}}}
    \caption{The variation of muon anomalous magnetic moment $\Delta a_{\mu}$ with coupling $y_{\psi}$ (\ref{gmzplane}(a)). The horizontal lines depict the experimental range of muon anomalous magnetic moment \cite{Muong-2:2021ojo}. The region of the parameter space in ($M_{\eta_{R}^{0}}-y_\psi$) plane which is in consonance with muon ($g-2$) and neutrino oscillation data (\ref{gmzplane}(b)).}
    \label{gmzplane}
  \end{figure}

The contribution to muon ($g-2$) is calculated using Eqn. (\ref{au2}). In Fig. \ref{gmzplane}(a), we have shown the correlation of muon anomalous magnetic moment $\Delta a_\mu$ with coupling constant $y_\psi$. Fig. \ref{gmzplane}(b) depicts the region of the parameter space in ($M_{\eta_{R}^{0}}-y_\psi$) plane which is consistent with observed range of $\Delta a_\mu$. The parameter space shown in Figs. \ref{gmzplane}(a) and \ref{gmzplane}(b) have been obtained by requiring the model to be consistent with neutrino oscillation data (Eqn. (\ref{ndata})). It is evident from Figs. \ref{gmzplane}(a) and \ref{gmzplane}(b) that the model explains muon anomalous magnetic moment $\Delta a_\mu$ for $1.3<y_\psi<2.8$ and mass of DM candidate in the range $152\text{ GeV}<M_{\eta_{R}^{0}}<195\text{ GeV}$, respectively.

\subsection*{W-boson Mass Anomaly }

\begin{figure} [t]%
    \centering
    {{\includegraphics[width=7cm]{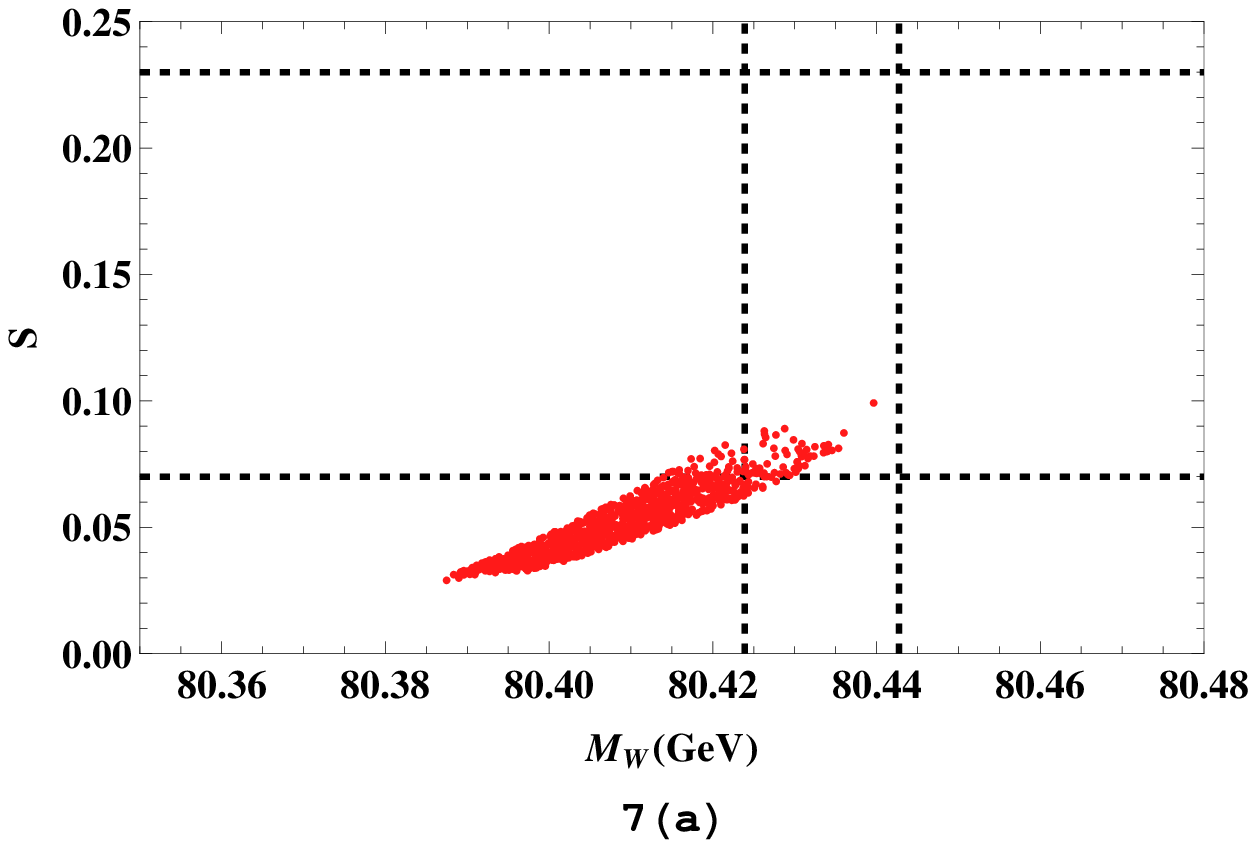}}}
    \qquad
    {{\includegraphics[width=7cm]{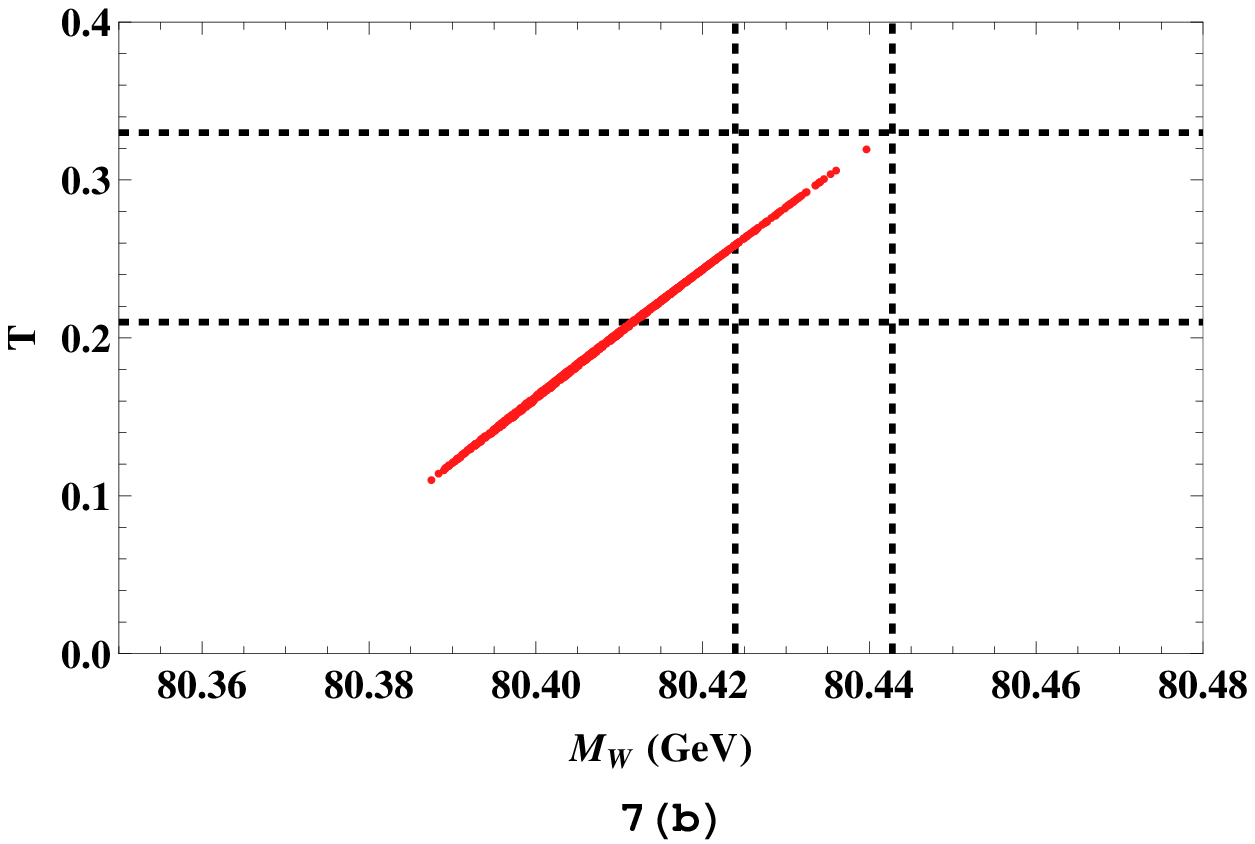}}}
    \caption{The correlation of oblique parameters $S$ and $T$ with W-boson mass $M_{W}$. The dashed lines indicate the experimental ranges of oblique parameters (horizontal) \cite{Lu:2022bgw} and CDF-II W-boson mass (vertical) \cite{CDF:2022hxs}.}%
   \label{stmw}%
\end{figure}

\begin{figure} [t]%
    \centering
    {{\includegraphics[width=7cm]{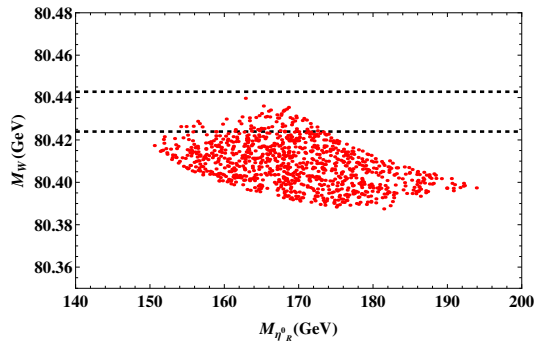}}}

    \caption{The correlation of mass of real part of inert doublet $M_{\eta_{R}^{0}}$ with W-boson mass $M_{W}$. The horizontal dashed lines indicate the CDF-II W-boson mass range\cite{CDF:2022hxs}. }%
    \label{mwri}%
\end{figure}

The correction to W-boson mass and oblique parameters reflecting the NP contribution are obtained using Eqns. (\ref{Mw}) and (\ref{T}), respectively. In Figs. \ref{stmw}(a) and \ref{stmw}(b), we have shown the correlations of oblique parameters $S$ and $T$ with W-boson mass $M_W$.  
The dashed lines depict experimental ranges of the respective parameters. The parameter space shown in Figs. \ref{stmw}(a) and \ref{stmw}(b) is, also,  consistent with the muon ($g-2$) and neutrino oscillation data. Further, it depicts the region accommodating CDF-II W-boson mass range and oblique parameters $S$ and $T$ predicted by the model.  Fig. (\ref{mwri}) shows the variation of W-boson mass $M_W$ with mass of real part of inert doublet $M_{\eta^{0}_{R}}$ (DM candidate). The dashed lines represent the experimental range of W-boson mass. It can be seen from  Fig. (\ref{mwri}) that the explanation of W-boson mass anomaly further constrain the mass of DM candidate, $M_{\eta_{R}^{0}}$, to be in the range $154\text{ GeV}<M_{\eta_{R}^{0}}<174\text{ GeV}$ in comparison of the range exhibited in Fig. \ref{gmzplane}(b) ($152\text{ GeV}<M_{\eta_{R}^{0}}<195\text{ GeV}$).

\subsection*{Dark Matter}
As discussed earlier, the real part of inert doublet is the lightest among all the BSM fields and is suitable DM candidate in the model. The mass of DM predicted by the model is in the range $154\text{ GeV}<M_{\eta_{R}^{0}}<174\text{ GeV}$ which is in consonance with neutrino oscillation data, muon ($g-2$) and explaining  W-boson mass anomaly, simultaneously.  The decay of DM into SM particles may be through two portals in the model \textit{viz.}, the Higgs portal and lepton portal via VLL triplet $\psi_T$.

\textbf{Benchmark Point:} For ready reference, we provide benchmark point for which the model satisfies neutrino oscillation data, muon ($g-2$) and W-boson mass anomaly, simultaneously. For input parameters as listed in Table \ref{tab3}, neutrino mixing angles and mass-squared differences predicted by the model are

\begin{equation*}
    \sin^{2} \theta_{13} = 0.022 ,\quad  \sin^{2} \theta_{23} = 0.60,\quad \sin^{2} \theta_{12} = 0.34 ,
    \end{equation*}
    \begin{equation*}
  \Delta m^{2}_{31} = 2.46\times 10^{-3} \text{eV}^{2} , \quad \Delta m^{2}_{21} = 7.1\times 10^{-5} \text{eV}^{2}.
    \end{equation*}
The value of $\Delta a_{\mu} =2.96\times 10^{-9} $ for $M_{\psi} = 269 $ GeV,  $M_{\eta} = 101 $ GeV and $y_{\psi} = 2.57$ while the mass of W-boson $M_{W}=80.43$ GeV for oblique parameters $S = 0.08$ and $T = 0.29$.

\begin{table}[t]
   
      \centering
        \begin{tabular}{ll}
             \hline
             \hline
       Parameters  &  Value \\
       \hline
       
        ($M_{11}$,  $M_{23}$) & (50.9, 68.4) GeV \\ 
          ($x, y$) &  (63.7, 25.5) GeV \\ 
         $ \delta^{\circ}$ & 333.4 \\ 
             $y_{\eta k}$  & (0.03, 0.05, 0.05)  \\
                ($M_{\eta_{R}^{0}}, M_{\eta_{I}^{0}}$) & (165.93, 165.94) GeV \\ 
                  $v_{S}$ & 116 GeV \\
                     $ M_{k}$ & (8.64, 2.30, 7) $\times 10^{7}$ GeV  \\ 
                   \hline
        \end{tabular}
   
    \caption{The values of input parameters used to obtain benchmark point of the numerical analysis ($k=1,2,3$).}
    \label{tab3}
\end{table}

Finally, a comment on Lepton flavour violation (LFV) process $\mu\rightarrow e\gamma$ is in order. The investigation of lepton flavour violation (LFV) process is encouraging from the perspective of explorable signatures of BSM scenarios at collider experiments. Any detection of such LFV decays like $\mu \rightarrow e \gamma$ will be a new physics signature. Within the model proposed in this work, the charged component of scalar doublet can provide such NP contribution for which the decay width is given by \cite{Toma:2013zsa,Lavoura:2003xp}
 
 \begin{eqnarray}
  \text{Br}(\mu \rightarrow e\gamma) = \frac{3(4\pi)^{3}\alpha _{em}}{4 G_{F}^2}|A_{D}|^{2} \text{Br}(\mu \rightarrow e\nu_{\mu}\bar{\nu_{e}}),
 \end{eqnarray}
 where 
 \begin{equation}
     A_{D} = \sum_{k}{} \frac{y^{*}_{ke} y^{}_{k\mu}}{16 \pi^{2}}\frac{1}{M_{\eta^{+}}^{2}} f(x'),
 \end{equation}
 $G_{F}$ is Fermi constant, $x' = M_{k}^{2}/M_{\eta^{+}}^{2}$ and $f(x')=\frac{1 - 6x' + 3x'^{2} +2x'^{3} - 6x'^{2}\log x'}{12(1-x')^{4}}$. The LFV prediction of the model is found to be in consonance with the experimental bound( in fact less than $10^{-14}$) for the obtained DM mass range.

\section{Conclusions}

Muon ($g-2$) anomaly, neutrino mass and mixing pattern, dark matter lack an explanation within the SM. Also,  recent observation of W-boson anomaly requires new physics explanation beyond SM. In this work, we propose a scotogenic scenario explaining neutrino oscillation data, muon ($g-2$), W-boson mass anomaly with a possible DM candidate, simultaneously. SM gauge group is extended with $Z_{4}$ symmetry wherein  a scalar singlet $S$ is introduced which acquires $vev$ and breaks this symmetry to the remnant $Z_{2}$  symmetry stabilizing the DM candidate in the model. In particular, to accommodate small neutrino masses and DM in the model, we have considered scotogenic model that includes three right-handed neutrinos and an inert scalar doublet which is odd under unbroken $Z_2$ symmetry. The real part of inert doublet is the lightest $Z_2$ odd particle, thus, it is a suitable DM candidate in the model. Furthermore, a VLL triplet $\psi_T$ is introduced which couples to scalar doublet providing positive contribution to muon ($g-2$). We, also, provide benchmark point of the numerical analysis indicating the simultaneous explanation of the above mentioned anomalies.   In addition, the implication of the model for $0\nu\beta\beta$ decay has, also, been studied. The model predicts a lower bound $m_{ee}>0.025$ eV at 3$\sigma$, which is well within the sensitivity reach of the  $0\nu\beta\beta$ decay experiments. The model, in general, predicts both $CP$ conserving and $CP$ violating solutions. The model explains muon anomalous magnetic moment $\Delta a_\mu$ for $1.3<y_\psi<2.8$ and mass of DM candidate in the range $152\text{ GeV}<M_{\eta_{R}^{0}}<195\text{ GeV}$ (Fig. \ref{gmzplane}(b)). The explanation of W-boson mass anomaly, further, constrain the mass of DM candidate, $M_{\eta_{R}^{0}}$, to be in the range $154\text{ GeV}<M_{\eta_{R}^{0}}<174\text{ GeV}$ (Fig. (\ref{mwri})).       
\vspace{1 cm}

\noindent \textbf{\Large{Acknowledgments}}\\
S. Arora acknowledges the financial support provided by the Central University of Himachal Pradesh. M. K. acknowledges the financial support provided by Department of Science and Technology, Government of India vide Grant No. DST/INSPIRE Fellowship/2018/IF180327. The authors, also, acknowledge Department of Physics and Astronomical Science for providing necessary facility to carry out this work. B. C. Chauhan is, also, thankful to the Inter University Centre for Astronomy and Astrophysics (IUCAA) for providing necessary facilities during the completion of this work.


\begin{thebibliography}{100}


\bibitem{Muong-2:2021ojo}
B.~Abi \textit{et al.} [Muon g-2],
Phys. Rev. Lett. \textbf{126}, no.14, 141801 (2021).
\bibitem{Aoyama:2020ynm}
T.~Aoyama \textit{et al.},
Phys. Rept. \textbf{887}, 1-166 (2020).

\bibitem{Abe:2017jqo}
T.~Abe, R.~Sato and K.~Yagyu,
JHEP \textbf{07}, 012 (2017).
\bibitem{Hutauruk:2020xtk}
P.~T.~P.~Hutauruk, D.~W.~Kang, J.~Kim and H.~Okada,
[arXiv:2012.11156 [hep-ph]].
\bibitem{Calibbi:2021qto}
L.~Calibbi, M.~L.~L\'opez-Ib\'a\~nez, A.~Melis and O.~Vives,
Eur. Phys. J. C \textbf{81}, no.10, 929 (2021).
\bibitem{Saez:2021qta}
B.~D.~S\'aez and K.~Ghorbani,
Phys. Lett. B \textbf{823}, 136750 (2021).





\bibitem{He:1990pn}
X.~G.~He, G.~C.~Joshi, H.~Lew and R.~R.~Volkas,
Phys. Rev. D \textbf{43}, 22-24 (1991).
\bibitem{He:1991qd}
X.~G.~He, G.~C.~Joshi, H.~Lew and R.~R.~Volkas,
Phys. Rev. D \textbf{44}, 2118-2132 (1991).













\bibitem{Borah:2021mri}
D.~Borah, A.~Dasgupta and D.~Mahanta,
Phys. Rev. D \textbf{104}, no.7, 075006 (2021).


\bibitem{Biswas:2016yan}
A.~Biswas, S.~Choubey and S.~Khan,
JHEP \textbf{09}, 147 (2016).
\bibitem{Zhou:2021vnf}
S.~Zhou,
Chin. Phys. C \textbf{46}, no.12, 011001 (2022).
\bibitem{Guo:2006qa}
W.~l.~Guo, Z.~z.~Xing and S.~Zhou,
Int. J. Mod. Phys. E \textbf{16}, 1-50 (2007).
\bibitem{Dev:2017fdz}
A.~Dev,
[arXiv:1710.02878 [hep-ph]].
\bibitem{Majumdar:2020xws}
C.~Majumdar, S.~Patra, P.~Pritimita, S.~Senapati and U.~A.~Yajnik,
JHEP \textbf{09}, 010 (2020).
\bibitem{Patra:2016shz}
S.~Patra, S.~Rao, N.~Sahoo and N.~Sahu,
Nucl. Phys. B \textbf{917}, 317-336 (2017).













\bibitem{ParticleDataGroup:2020ssz}
P.~A.~Zyla \textit{et al.} [Particle Data Group],
PTEP \textbf{2020}, no.8, 083C01 (2020).
\bibitem{CDF:2022hxs}
T.~Aaltonen \textit{et al.} [CDF],
Science \textbf{376}, no.6589, 170-176 (2022).

\bibitem{CentellesChulia:2022vpz}
S.~Centelles Chuli\'a, R.~Srivastava and S.~Yadav,
[arXiv:2206.11903 [hep-ph]].
\bibitem{YaserAyazi:2022tbn}
S.~Yaser Ayazi and M.~Hosseini,
[arXiv:2206.11041 [hep-ph]].
\bibitem{Afonin:2022cbi}
S.~S.~Afonin,
[arXiv:2205.12237 [hep-ph]].
\bibitem{Kawamura:2022fhm}
J.~Kawamura and S.~Raby,
[arXiv:2205.10480 [hep-ph]].
\bibitem{Wang:2022dte}
J.~W.~Wang, X.~J.~Bi, P.~F.~Yin and Z.~H.~Yu,
[arXiv:2205.00783 [hep-ph]].
\bibitem{Batra:2022pej}
A.~Batra, S.~K.~A, S.~Mandal, H.~Prajapati and R.~Srivastava,
[arXiv:2204.11945 [hep-ph]].
\bibitem{Borah:2022zim}
D.~Borah, S.~Mahapatra and N.~Sahu,
Phys. Lett. B \textbf{831}, 137196 (2022).
\bibitem{Ghorbani:2022vtv}
K.~Ghorbani and P.~Ghorbani,
[arXiv:2204.09001 [hep-ph]].
\bibitem{Borah:2022obi}
D.~Borah, S.~Mahapatra, D.~Nanda and N.~Sahu,
[arXiv:2204.08266 [hep-ph]].
\bibitem{Popov:2022ldh}
O.~Popov and R.~Srivastava,
[arXiv:2204.08568 [hep-ph]].
\bibitem{Chakrabarty:2022voz}
N.~Chakrabarty,
[arXiv:2206.11771 [hep-ph]].
\bibitem{Chowdhury:2022dps}
T.~A.~Chowdhury and S.~Saad,
[arXiv:2205.03917 [hep-ph]].
\bibitem{He:2022zjz}
S.~P.~He,
[arXiv:2205.02088 [hep-ph]].
\bibitem{Dcruz:2022dao}
R.~Dcruz and A.~Thapa,
[arXiv:2205.02217 [hep-ph]].
\bibitem{Kim:2022xuo}
J.~Kim,
Phys. Lett. B \textbf{832}, 137220 (2022).
\bibitem{Kim:2022hvh}
J.~Kim, S.~Lee, P.~Sanyal and J.~Song,
[arXiv:2205.01701 [hep-ph]].
\bibitem{Botella:2022rte}
F.~J.~Botella, F.~Cornet-Gomez, C.~Mir\'o and M.~Nebot,
[arXiv:2205.01115 [hep-ph]].
\bibitem{Zhou:2022cql}
Q.~Zhou and X.~F.~Han,
[arXiv:2204.13027 [hep-ph]].
\bibitem{Baek:2022agi}
S.~Baek,
[arXiv:2204.09585 [hep-ph]].
\bibitem{Bhaskar:2022vgk}
A.~Bhaskar, A.~A.~Madathil, T.~Mandal and S.~Mitra,
[arXiv:2204.09031 [hep-ph]].
\bibitem{Han:2022juu}
X.~F.~Han, F.~Wang, L.~Wang, J.~M.~Yang and Y.~Zhang,
doi:10.1088/1674-1137/ac7c63
[arXiv:2204.06505 [hep-ph]].
\bibitem{Borah:2021khc}
D.~Borah, M.~Dutta, S.~Mahapatra and N.~Sahu,
Phys. Rev. D \textbf{105}, no.1, 015029 (2022).
\bibitem{Jana:2020joi}
S.~Jana, P.~K.~Vishnu, W.~Rodejohann and S.~Saad,
Phys. Rev. D \textbf{102}, no.7, 075003 (2020).
\bibitem{Baek:2015fea}
S.~Baek,
Phys. Lett. B \textbf{756}, 1-5 (2016).
\bibitem{Han:2019diw}
Z.~L.~Han, R.~Ding, S.~J.~Lin and B.~Zhu,
Eur. Phys. J. C \textbf{79}, no.12, 1007 (2019)
\bibitem{Kang:2021jmi}
D.~W.~Kang, J.~Kim and H.~Okada,
Phys. Lett. B \textbf{822}, 136666 (2021)


\bibitem{Ma:2006km}
E.~Ma,
Phys. Rev. D \textbf{73}, 077301 (2006).
\bibitem{Merle:2015ica}
A.~Merle and M.~Platscher,
JHEP \textbf{11}, 148 (2015).

\bibitem{Freitas:2014pua}
A.~Freitas, J.~Lykken, S.~Kell and S.~Westhoff,
JHEP \textbf{05}, 145 (2014)
[erratum: JHEP \textbf{09}, 155 (2014)].











\bibitem{Peskin:1991sw}
M.~E.~Peskin and T.~Takeuchi,
Phys. Rev. D \textbf{46}, 381-409 (1992).

\bibitem{Lu:2022bgw}
C.~T.~Lu, L.~Wu, Y.~Wu and B.~Zhu,
[arXiv:2204.03796 [hep-ph]].
\bibitem{Zhang:2006vt}
H.~H.~Zhang, W.~B.~Yan and X.~S.~Li,
Mod. Phys. Lett. A \textbf{23}, 637-646 (2008).

\bibitem{Esteban:2020cvm}
I.~Esteban, M.~C.~Gonzalez-Garcia, M.~Maltoni, T.~Schwetz and A.~Zhou,
JHEP \textbf{09}, 178 (2020).



\bibitem{Krastev:1988yu}
P.~I.~Krastev and S.~T.~Petcov,
Phys. Lett. B \textbf{205}, 84-92 (1988).
\bibitem{Jarlskog:1985ht}
C.~Jarlskog,
Phys. Rev. Lett. \textbf{55}, 1039 (1985).
\bibitem{Licciardi:2017oqg}
C.~Licciardi [nEXO],
J. Phys. Conf. Ser. \textbf{888}, no.1, 012237 (2017).
\bibitem{NEXT:2009vsd}
F.~Granena \textit{et al.} [NEXT],
[arXiv:0907.4054 [hep-ex]].
\bibitem{NEXT:2013wsz}
J.~J.~Gomez-Cadenas \textit{et al.} [NEXT],
Adv. High Energy Phys. \textbf{2014}, 907067 (2014).
\bibitem{KamLAND-Zen:2016pfg}
A.~Gando \textit{et al.} [KamLAND-Zen],
Phys. Rev. Lett. \textbf{117}, no.8, 082503 (2016).
\bibitem{Barabash:2011row}
A.~S.~Barabash,
J. Phys. Conf. Ser. \textbf{375}, 042012 (2012).




\bibitem{Krnjaic:2019rsv}
G.~Krnjaic, G.~Marques-Tavares, D.~Redigolo and K.~Tobioka,
Phys. Rev. Lett. \textbf{124}, no.4, 041802 (2020).

\bibitem{Gninenko:2014pea}
S.~N.~Gninenko, N.~V.~Krasnikov and V.~A.~Matveev,
Phys. Rev. D \textbf{91}, 095015 (2015).
\bibitem{Bauer:2018onh}
M.~Bauer, P.~Foldenauer and J.~Jaeckel,
JHEP \textbf{07}, 094 (2018).
\bibitem{Kamada:2018zxi}
A.~Kamada, K.~Kaneta, K.~Yanagi and H.~B.~Yu,
JHEP \textbf{06}, 117 (2018).

\bibitem{Giusarma:2016phn}
E.~Giusarma, M.~Gerbino, O.~Mena, S.~Vagnozzi, S.~Ho and K.~Freese,
Phys. Rev. D \textbf{94}, no.8, 083522 (2016).
\bibitem{Planck:2018vyg}
N.~Aghanim \textit{et al.} [Planck],
Astron. Astrophys. \textbf{641}, A6 (2020)
[erratum: Astron. Astrophys. \textbf{652}, C4 (2021)].


\bibitem{Toma:2013zsa}
T.~Toma and A.~Vicente,
JHEP \textbf{01}, 160 (2014).
\bibitem{Lavoura:2003xp}
L.~Lavoura,
Eur. Phys. J. C \textbf{29}, 191-195 (2003).


\end{thebibliography}
\end{document}